\documentclass[12pt,a4paper]{article}
\usepackage[]{amsmath,amssymb}
\usepackage{graphics,epsfig}
\usepackage{graphicx, color}
\usepackage{geometry}

\begin{document}

\date{}
\title{On the RG running of the entanglement entropy of a circle}
\author{H. Casini\footnote{e-mail: casini@cab.cnea.gov.ar} 
 \, and M. Huerta\footnote{e-mail: marina.huerta@cab.cnea.gov.ar} \\
{\sl Centro At\'omico Bariloche,
8400-S.C. de Bariloche, R\'{\i}o Negro, Argentina}}\maketitle

\begin{abstract}
We show, using strong subadditivity and Lorentz covariance, that in three dimensional space-time the entanglement entropy of a circle is a concave function. This implies the decrease of the coefficient of the area term and the increase of the constant term in the entropy between the ultraviolet and infrared fixed points. This is in accordance with recent holographic c-theorems and with conjectures about the renormalization group flow of the partition function of a three sphere (F-theorem). 
 The irreversibility of the renormalization group flow in three dimensions would follow from the argument provided there is an intrinsic definition  for the constant term in the entropy at fixed points. We discuss the difficulties in generalizing this result for spheres in higher dimensions.  
\end{abstract}

\section{Introduction}

The Zamolodchikov's c-theorem in two dimensions shows the renormalization group flow can only connect conformal field theories (CFT) at the ultraviolet (UV) fixed point with higher Virasoro central charge than the CFT at the infrared (IR) fixed point \cite{zamo}. This establishes an ordering on the CFT in two dimensions, providing an interpretation of the central charge as a measure of the field degrees of freedom. These degrees of freedom are lost along renormalization flow due to the decoupling of massive modes. 
Zamolodchikov's proof is based on positivity and covariance of the correlation functions. An alternative proof of the same theorem based on the strong subadditivity of the entanglement entropy was given in \cite{doscteo}. In both cases the essential elements behind the validity of the theorem are unitarity and relativistic symmetry of quantum field theory (QFT). 

Along the years much work and progress has been made to find different c-functions with monotonic behavior under the renormalization group (RG) flow in higher dimensions. An interesting proposal for even space-time dimensions is to identify the monotonous quantity with the coefficient of the Euler term in the trace anomaly \cite{anomaly}. This has received much support from explicit examples \cite{examples} as well as from investigations based on general properties of QFT \cite{invest}. This anomaly coefficient has also an expression in terms of the entanglement entropy as the coefficient of the logarithmic term in the entropy of spheres in even dimensions \cite{solo,n-sphere,otrosholo,hhm}.
   
Recently, several c-theorems where obtained holographicaly \cite{otrosholo,holoholo,aninda}. For even dimensions they give support to the monotonic running of the above mentioned anomaly coefficient. See \cite{qq} for further developments. For odd space-time dimensions, where the trace anomaly is absent, Myers and Sinha  proposed the constant term in the entanglement entropy of a sphere as the relevant c-function \cite{aninda} (see also \cite{otrosholo,masivocurva}). They find this term changes monotonically along the renormalization group flow in holographic theories by connecting this property to the null energy condition in the AdS-CFT context. 

In an apparently unrelated work, it was recently discovered that the free energy $F=-\log Z$ on a three sphere in certain supersymmetric theories decreases under the renormalization group transformations \cite{ff}. This property was called F-theorem, and conjectured to be valid for any three dimensional theory. This conjecture has been further checked in several models \cite{ff1}. The partition function on the three sphere and the entanglement entropy of the circle for conformal theories where shown to be proportional in \cite{hhm} (see also \cite{doww}), thus establishing a connection between the F-theorem and the monotonous running of the finite part of the circle entanglement entropy.

In this paper we extend the analysis of \cite{doscteo} to the case of circles in $d=2$ spatial dimensions. We consider the combined constraints imposed by SSA and Lorentz invariance to the entropy. In contrast to the one dimensional case, here we need to consider the SSA inequalities involving an arbitrarily large number of circles. This is necessary in order to obtain circles at both sides of the inequalities in the limit of large number of regions. We find the entropy of circles in $2+1$ dimensions have a concave entropy. This implies the constant term increases from the UV to the IR fixed points. In more dimensions a better understanding of the structure of divergent terms for non-smooth entangling surfaces is necessary to generalize the three dimensional construction.

The plan of the paper is as follows. In the next section we review the entropic form of the c-theorem in $1+1$ dimensions. In section 3 we introduce an inequality for the entanglement entropy of an arbitrary number of regions which follows from SSA. In section 4 we show the entanglement entropy of circles in $2+1$ dimensions is a concave function of the radius. An attempt to  generalize this calculation to any dimensions is given in section 5. Finally, in section 6 we end with some discussion.     
   
\section{Intervals in two-dimensional space-time}

The strong subadditive inequality (SSA) for the entanglement entropy of spatial regions $A$ and $B$ write 
\begin{equation}
S(A)+S(B)\geq S(A\cap B)+S(A\cup B) \,.
\end{equation}

\begin{figure}
\centering
\leavevmode
\epsfysize=4cm
\epsfbox{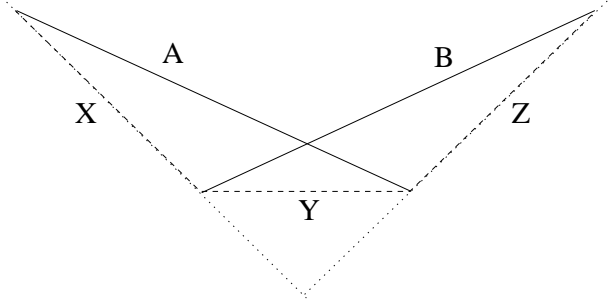}
\bigskip
\caption{Two interval $A$ and $B$ of size $\sqrt{r R}$ have their end-points on the light-cone. The size of $Y$ is $r$ while the distance between the upper end-points of $A$ and $B$ is $R$.}
\label{f3}
\end{figure}

Consider two boosted intervals $A$ and $B$ in $1+1$ dimensions with end-points located on a light cone as shown in figure \ref{f3}. Applying SSA to the spatial regions in the dashed line in figure \ref{f3} we have 
\begin{equation}
S(XY)+S(YZ)\ge S(Y)+S(XYZ)\,.\label{tito}
\end{equation}    
For convenience, we choose the size of $A$ and $B$ to be $\sqrt{r R}$, for some $r<R$, and the size of $Y$ equal to $r$, $S(Y)=S(r)$. We are writing $S(l)$ for the entanglement entropy of an interval of length $l$. Since by causality the density matrix of a region coincides with the one of any other spatial region with the same causal domain of dependence, we have $S(XY)=S(A)=S(\sqrt{r R})=S(B)=S(YZ)$. Also, the region $XYZ$ is equivalent to an interval of size $R$, hence $S(XYZ)=S(R)$. Using this in (\ref{tito}) leads to 
\begin{equation}
2 \, S(\sqrt{rR})\ge S(R)+S(r)\,.
\end{equation}
Setting $R=r+\varepsilon$ and expanding for small $\varepsilon$ we get 
\begin{equation}
r S^{^{\prime \prime }}(r)+S^{^{\prime }}(r)\leq 0\,.
\end{equation}
This is equivalent to $C^{\prime}(r)\le 0$ for the quantity $C(r)=r S^{^{\prime }}(r)$. This c-function is then dimensionless and always decreasing.   
 At the critical point the entropy has a general form  
\begin{equation}
S(r)=\frac{c}{3}\log (r/\epsilon)+c_0\,, 
\end{equation}
where $c_0$ is a non-universal constant, $\epsilon$ is a short distance cutoff, and $c$ the Virasoro central charge of the conformal theory. In this case we have $C(r)=c/3$. Therefore the above inequality establishes an entropic form of the 
Zamolodchikov's c-theorem \cite{doscteo,nues2}: the central charge of the ultraviolet fixed point is always greater than the one of the infrared fixed point.

Let us isolate $c_0$ form the entropy in order to see how it runs with the renormalization group. Define 
\begin{equation}
c_0(r)=S(r)-r\,\log (r/\epsilon)\,S^{^{\prime }}(r) \,,
\end{equation}
which coincides with $c_0$ at the conformal point. We have 
\begin{equation}
c_0^{^{\prime }}(r)=-\log (r/\epsilon)\,(r\,S^{^{\prime }}(r))^{^{\prime }}=-\log
(r/\epsilon)\,C^{^{\prime }}(r) \,.
\end{equation}
Integrating between the UV and IR fixed points we have
\begin{equation}
\Delta c_0=c_0^{\textrm{ir}}-c_0^{\textrm{uv}}=-\int_0^\infty dr\,\log (r/ \epsilon)\,C^{^{\prime }}(r) \ge 0\,.\label{ccc}
\end{equation}
This is always positive, and $c_0$ at the infrared is greater than $c_0$ the ultraviolet point. However, the constant $c_0$ and its total variation are not universal quantities and depend on the cutoff. For example, for a massive field at the infrared the entropy is a divergent constant $S\sim c_0=\frac{c}{3}\log (m/\epsilon
)$ \cite{logmasa} ($\Delta c_0$ can also have infrared divergences for scalar fields).

According to (\ref{ccc}) the only way $\Delta c_0$ is finite is that $C^{\prime}=0$, that is, the IR and UV conformal theories have  the same central charge. But in this case there is no running at all in a relativistic theory, and $\Delta c_0=0$, as shown by (\ref{ccc}).  However, the situation changes when a boundary and boundary conditions are imposed on the theory. For a fixed bulk conformal theory the boundary conditions introduce a boundary contribution $c_0=\log (g)$ to the entanglement entropy, where $\log (g)$ is called the boundary entropy \cite{logmasa}. The boundary conditions run with the renormalization group producing a  finite running in $\log (g)$. It is known that $\log (g)$ decreases to the infrared under the renormalization group \cite{gteo}. This is called  the $g$-theorem. Thus, this boundary induced running for $c_0$ is opposite to the one above, related to the running of the bulk theory (see \cite{analogous} for a related conclusion). Our argument above for $\Delta c_0>0$ of course does not apply directly when there are boundary conditions breaking boost symmetries, though it would be interesting if a connection with the $g$-theorem could be established (see \cite{ryuu} for a related argument).  

\section{Symmetric form of SSA for many subsystems}
There are several obstacles to generalize the above construction for $d\ge 2$ spatial dimensions \cite{nues2}. The most evident one is that in general the intersection and union of regions of a given shape are not of the same shape in more dimensions, specially if these regions are boosted to each other. 

It is possible to avoid this problem in some specific cases by taking the limit of an infinite number of regions whose intersections and union tend to the desired shape in the limit. 

For three sets we have repeatedly using SSA 
\begin{eqnarray}
S(A)+S(B)+S(C) &\geq &S(A\cap B)+S(A\cup B)+S(C) \\
&\geq &S(A\cup B\cup C)+S((A\cup B)\cap C)+S(A\cap B)\nonumber \\
&\geq &S(A\cup B\cup C)+S(((A\cup B)\cap C)\cup (A\cap B))+S(A\cap B\cap C)
\nonumber \\
&=&S(A\cup B\cup C)+S((A\cap C)\cup (A\cap B)\cup (B\cap C))+S(A\cap B\cap C)\nonumber \,.
\end{eqnarray}
 Using this same idea for an arbitrary number of sets $X_{i}$, with $i=1,2,...,N$, one shows by induction that  
\begin{equation}
\sum_{i}S(X_{i}) \geq S(\cup _{i}X_{i})+S(\cup _{\{ij\}}(X_{i}\cap
X_{j}))+S(\cup _{\{ijk\}}(X_{i}\cap X_{j}\cap X_{k}))+...   +S(\cap _{i}X_{i}) \,. \label{ecu}
\end{equation}
There are $N$ terms on each side. The sets on the right hand side are ordered by inclusion from right to left, and each one of the terms is totally symmetrical under permutations of the $N$ regions $X_i$.

\section{Circles in 2+1 dimensions}

\begin{figure}
\centering
\leavevmode
\epsfysize=6cm
\epsfbox{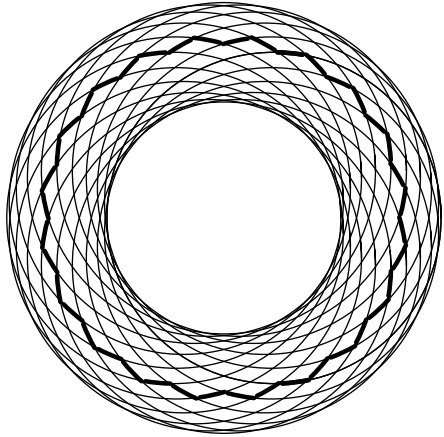}
\bigskip
\caption{Circles on a plane obtained by rotating a single circle around a point different from its center. The marked contour shows a typical set involved in the series of the right hand side of (\ref{ecu}). This set approaches a circle of specific radius as the number $N$ of rotated circles increases, but the perimeter of the resulting shape is greater than the one corresponding to a circle. Also, since the slope of the wiggles around the limit circle does not change with $N$, the contribution of the corners to the entropy does not vanish in the large $N$ limit.}
\label{f1}
\end{figure}

We want to use (\ref{ecu}) for objects whose intersection and union do not belong to the same class of the original sets, but when a limit of infinitely many sets is taken these do belong to the same class. As an example, the intersection and union of circles in the plane gives place to different shapes as shown in figure \ref{f1}. But using $N$ copies of circles rotated an angle  $\frac{2\pi k}{N}$ with $k=1,2,..N$ around a point different from the center, and taking the infinite $N$ limit, the sets at the right hand side of (\ref{ecu}) look like circles centered at the same point. Of course, we have to see if the limit of the entropies goes smoothly to that of the circles. In the case of circles in the plane the answer is no, since for any intersection and union of a pair of circles the SSA relation gives place to an inequality with infinite difference between both sides. This is due to the presence of corners in the intersection and union of circles. The corners give place to logarithmically divergent contributions on the entropy which are not present in the circles \cite{corners}. Also, the regions in the right hand side of (\ref{ecu}) approach circles (in some unspecified ``eye-sight'' topology) but their perimeters do not approach the one of the corresponding circle (see figure \ref{f1}), and the divergent parts of the entropies do not match. Therefore the inequality is trivially satisfied and no universal information can be obtained from this construction.

However, if we take boosted circles which have the boundary in the light cone $t^2=x^2+y^2$, even if these wiggles for the sets in the right hand side of (\ref{ecu}) are still there, they are located on a null surface. This has two important consequences. First, the perimeter of the regions coincides with that of smooth circles. Second, the angles of the corners are equal to $\pi$. In fact, one can see that a corner in a null plane can be approached by a corner on a boosted spatial plane, in the large boost limit, provided the  corner angle (as seen in the reference system at rest with respect to the spatial plane) tends to $\pi$. For a corner in a spatial plane, the logarithmically divergent contribution to the entropy is a function of the corner angle $\pi-\delta$ which vanishes quadratically in $\delta$ for $\delta$ tending to zero\footnote{It is possible to interpolate smoothly between the case of $N$ circles on the plane in figure \ref{f1} and the circles on the null cone by drawing boosted circles on a spatial hyperboloid $t^2-x^2-y^2=\Lambda^2$. As the curvature of the hyperboloid goes to infinity ($\Lambda\rightarrow 0$), approaching the null cone, all the corner angles tend to $\pi$.}. Hence, we expect the entropy of the wiggly regions really approaches the one of a circle in this case. 

\begin{figure}
\centering
\leavevmode
\epsfysize=5cm
\epsfbox{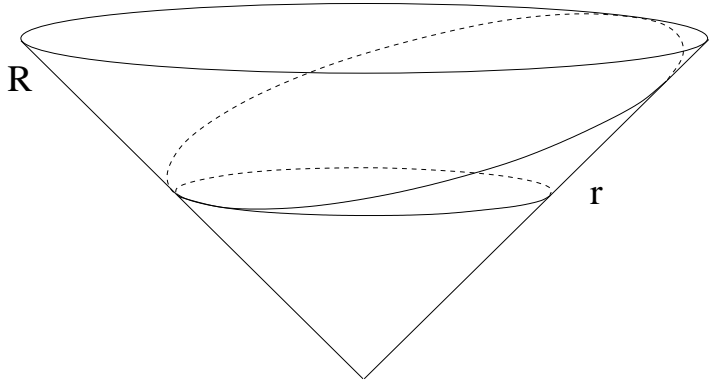}
\bigskip
\caption{The figure shows two parallel circles of radius $r$ and $R$ obtained by cutting the light-cone with two parallel spatial planes. Another spatial plane cuts the light-cone in a circle of radius $\sqrt{r R}$ which is tangent to the two parallel circles.}
\label{f2}
\end{figure}

Then, consider the intersection of a spatial plane and the null cone, which determines a (boosted) circle $D$ (see figure \ref{f2}). We obtain a series of $N$  of these boosted circles, $D_k$, by rotating the $x,y$ coordinates the angles $\frac{2 \pi k}{N}$, $k=0,...,N-1$. The view of this construction projected on the $x,y$ plane is similar to figure (\ref{f1}), but where the circles are replaced by ellipses. The SSA relation requires the spatial regions involved to have their boundaries spatial to each other \cite{geo}. We do not have to worry about this requirements since all the boundaries are located on the null cone. Also, we do not need to keep track of the surface itself, since the boundary determines the class of all spatial surfaces with equivalent domain of dependence.  

Let $R$ be the radius of the circle which is formed by the union of the rotated boosted circles in the infinite $N$ limit, and $r$ the one of the intersection of all of them.
The radius of the circle $D$ is then given by $\sqrt{Rr}$. Taking $D$ on the plane  
\begin{equation}
t=x\frac{(R-r)}{(R+r)}+\frac{2 r R}{r+R}\,,  \label{plano}
\end{equation}
the circle center is the point 
\begin{equation}
(x,y,t)=\left(\frac{R-r}{2},0,\frac{R+r}{2}\right)\,. 
\end{equation}
The circle equation is then given by 
\begin{equation}
Rr=\left( x-\frac{R-r}{2}\right) ^{2}+y^{2}-\left( t-\frac{R+r}{2}\right)
^{2} \,, \label{circulo}
\end{equation}
supplemented by the plane equation (\ref{plano}). 

The different radius $l$ for the approximate circles which appear in the right hand side of (\ref{ecu}) are given by the distances to the origin of the projections to the $x, y$ plane of the  intersections of $D$ with the rotated copies of $D$.
For a rotation angle $\theta$ this intersection point has
\begin{equation}
y=x\tan \left( \frac{\theta }{2}\right) \,.
\end{equation}
Then, using (\ref{plano}) and (\ref{circulo}) we get 
\begin{equation}
l=\sqrt{x^{2}+y^{2}}=\frac{2rR}{R+r-(R-r)\cos (\frac{\theta }{2})} \,.
\end{equation}
Therefore, the inequality (\ref{ecu}) reads 
\begin{equation}
N\,S(\sqrt{Rr})\geq \sum_{i=1}^{N}\tilde{S}\left( \frac{2rR}{R+r-(R-r)\cos (
\frac{\pi i}{N})}\right)\,, 
\end{equation}
where we have written $\tilde{S}$ instead of $S$ in order to remark that these are circle entropies only approximately. The limit $N\rightarrow \infty$ becomes 
\begin{equation}
S(\sqrt{Rr})\geq \frac{1}{\pi }\int_{0}^{\pi }dz\,S\left( \frac{2rR}{
R+r-(R-r)\cos (z)}\right)=\frac{1}{\pi }\int_{r}^{R}dl\,\frac{\sqrt{Rr}}{l\sqrt{
(R-l)(l-r)}}S(l)  \,. \label{ssai}
\end{equation}

For a term in the entropy proportional to the circle perimeter this inequality turns out to be an equation, that is, we have 
\begin{equation}
\frac{1}{\pi }\int_{0}^{\pi }dz\,\frac{2rR}{R+r-(R-r)\cos (z)}=\sqrt{Rr} \,.
\end{equation}
The equation also holds for a constant term. We could have anticipated the equality in these cases because the SSA inequality (for two regions) is already an equation for the area and the constant terms (this latter for intersecting regions).  
Writing $R=r+\varepsilon $ and expanding (\ref{ssai}) for small $
\varepsilon $   produces 
\begin{equation}
S^{^{\prime \prime }}\leq 0 \,.\label{infinit}
\end{equation}

At the conformal point we expect the entropy of the circle to be a linear function of the radius (for spheres the logarithmic term is related to the conformal anomaly, and this is absent for odd space-time dimensions \cite{hhm})
\begin{equation}
S(r)=c_{0}+c_{1}r \,,
\end{equation}
and consequently $S^{^{\prime \prime }}=0$. The total renormalization of the $c_1$ coefficient between fixed points reads
\begin{equation}
\Delta c_{1}=c_{1}^{\text{uv}}-c_{1}^{\text{ir}}=S^{^{\prime }\text{uv}
}-S^{^{\prime }\text{ir}}=-\int_{0}^{\infty }dr\, S^{^{\prime \prime }}\geq 0 \,.\label{c11}
\end{equation}

Since the second derivative of the entropy eliminates the divergent term, we expect that $S^{\prime \prime}(r)$ is universal, and the same applies to $c_{1}^{\text{uv}}-c_{1}^{\text{ir}}$.  In the free case the total running of $c_1$ is $\Delta c_1=\frac{\pi}{6} m $, both for massive scalars \cite{massiveterm} and Dirac fermions \cite{massiveterm1,kaba}. 

Now let us turn attention to the running of $c_0$.  Notice that the function $S(r)$ is concave and it is defined for $r>0$. Then the height at the origin of the asymptotic tangent at large $r$ has to be greater than the one of the tangent at $r=0$. To be more precise, consider the quantity 
\begin{equation}
c_0(r)=S(r)-r \,S^{\prime}(r)\,,\label{funi}
\end{equation}
 which coincides with $c_0$ at the fixed points. We have 
\begin{equation}
c_0^{\prime}(r)=-r \, S^{\prime\prime}(r)\ge 0\,.\label{ghgh}
\end{equation}
Then,
\begin{equation}
\Delta c_0=c_0^{\textrm{ir}}-c_0^{\textrm{uv}}=-\int_0^\infty dr\,r \, S^{\prime\prime}(r)\ge 0\,.\label{ty}
\end{equation}
In a recent paper \cite{masivocurva}, while searching for an adequate c-function corresponding to the running of the constant term, the authors define this same function (\ref{funi}) and conjecture eq. (\ref{ghgh}).

Because of $S^{\prime\prime}(r)$ should be universal, the same is expected for the quantity in (\ref{ty}). Thus, $\Delta c_0$ is  well defined.  We also expect it is finite in general, since at large mass the entropy would have an expansion of the form $S(r)\sim (k_1+ k_2 m ) r+ c_0 + \frac{k3}{m r}+\frac{k4}{(m r)^2}+...$. This is the case for free fields \cite{massiveterm,massiveterm1}. This is also the case in generic holographic models \cite{masivocurva}. Then $S^{\prime\prime}(r)\sim r^{-3}$ and the integral (\ref{ty}) is convergent at $r=\infty$.

The relation $\Delta c_1\ge 0$ cannot be consider an analogous to the c-theorem. Even disregarding the question of the universality for the moment, since $c_1$ is dimensionfull, it cannot depend on the CFT alone, and keeps track of the path in the renormalization trajectory from the UV to the IR fixed points. For example, a massive free field has a massless free UV fixed point and no degree of freedom at the IR. However, $\Delta c_1=\frac{\pi}{6} m$ depends on the specific mass of the field connecting these limits.

 On the other hand, $c_0$ is dimensionless. The relation $\Delta c_0\ge 0$ was first obtained holographicaly and proposed as a generalization of the c-theorem to three dimensions in \cite{aninda}. As mentioned in the introduction, this property is equivalent to the F-theorem for the three sphere partition functions \cite{ff}. However, we want to note here that these results, at least when referring to the entanglement entropy, could be of a different nature than the two dimensional c-theorem. They establish the increase (decrease) of a quantity from the UV to the IR fixed points, and the total change is well defined. However, the quantity itself at the fixed point is not. At present it is not possible to establish a correspondence of $c_0$ with some physical property of the continuum theory. There is an indefinition of $c_0$ by an additive constant which we do not know how to resolve. This is related to the fact that in the expression $S(r)=c_1 r+c_0$ the constant $c_1=\textrm{const}/\epsilon$, with $\epsilon$ a distance cutoff. Because there is a cutoff $\epsilon$, we cannot fix the physical size of the radius $r$ better than an error of order $\epsilon$. Changing the definition of the radius $r\rightarrow r+\textrm{const} \,\epsilon$ we change $c_0$ at will. This is clear for example in lattice evaluations of the free entanglement entropy of a sphere using the radial discretization of the Srednicki method \cite{sred}. In order to compute the entropy function is necessary to choose a definition of what the radius means in the lattice in terms of a number times the lattice spacing. While variations of this choice do not change the continuum limit of the correlation functions (and hence they all give place to the same continuum theory) they do alter the definition of $c_0$. A possible way out of this indetermination is to fix the global constant such that it takes the value zero for the limit of no degrees of freedom (by adding masses for all the fields)\footnote{In \cite{masivocurva} the authors computed $c_0(r)=r S^{\prime}(r)-S(r)$ numerically for a free massive scalar and find evidence for universality as well as for $c_0$ going to zero at the IR. However, we think this is tied to the special choice $r=(n+1/2)\epsilon$ they make for the radius in terms of the number $n$ of lattice points.}. This allows to compute a unique $c_0$ for any theory and specific way to go to zero degree of freedom. But it does not rule out the possibility that the resulting $c_0$ could be RG path dependent. If this is the case, the running (\ref{ty}) cannot be used to compare conformal theories at the fixed points. In particular, it would not rule out cycles in the renormalization group.

\bigskip

Perhaps it is of interest to end this section showing that the finite inequality (\ref{ssai}) contains the same information of the infinitesimal one (\ref{infinit}). Write $u=
\sqrt{Rr}$, $v=\sqrt{\frac{R}{r}}>1$, and (\ref{ssai}) becomes
\begin{equation}
S(u)\geq \frac{1}{\pi }\int_{0}^{\pi }dz\,S\left( g(u,v,z)\right) =h(u,v)\,,\label{utu}
\end{equation}
with 
\begin{equation}
g(u,v,z)=\frac{2u}{v+\frac{1}{v}-(v-\frac{1}{v})\cos (z)} \,.
\end{equation}
The variation of $h(u,v)$ with respect to $v$ is
\begin{equation}
\frac{\partial h(u,v)}{\partial v}=\frac{1}{\pi }\int_{0}^{\pi
}dz\,S^{^{\prime }}\left( g(u,v,z)\right) \frac{\partial g(u,v,z)}{\partial v
}\,, 
\end{equation}
with  
\begin{equation}
\frac{\partial g(u,v,z)}{\partial v}=\frac{(-2u)\left( 1-\frac{1}{v^{2}}-(1+
\frac{1}{v^{2}})\cos (z)\right) }{\left( v+\frac{1}{v}-(v-\frac{1}{v})\cos
(z)\right) ^{2}} \,. \label{gv}
\end{equation}
Now we write
\begin{equation}
\frac{\partial h(u,v)}{\partial v}=\frac{1}{\pi }\int_{0}^{\pi
}dz\,S^{^{\prime }}\left( g(u,v,z)\right) \frac{\partial }{\partial z}
f(u,v,z) \,,
\end{equation}
with 
\begin{equation}
f(u,v,z)=\int_{0}^{z}dz^{^{\prime }}\frac{\partial g(u,v,z^{^{\prime }})}{
\partial v} =\frac{2 u \sin(z)}{1+v^2+(1-v^2) \cos(z)}\,.\label{gv2}
\end{equation}
Then 
\begin{equation}
\frac{\partial h(u,v)}{\partial v} =\frac{1}{\pi }\left| S^{^{\prime
}}\left( g(u,v,z)\right) f(u,v,z)\right| _{z=0}^{z=\pi }-  
\frac{1}{\pi }\int_{0}^{\pi }dz\,S^{^{\prime \prime }}\left(
g(u,v,z)\right) \frac{\partial g(u,v,z)}{\partial z}f(u,v,z)\,.
\end{equation}
By explicit evaluation $f(u,v,0)=f(u,v,\pi)=0$. Then
\begin{equation}
\frac{\partial h(u,v)}{\partial v}=-\frac{1}{\pi }\int_{0}^{\pi
}dz\,S^{^{\prime \prime }}\left( g(u,v,z)\right) \frac{\partial g(u,v,z)}{
\partial z}f(u,v,z) \,.
\end{equation}
We have seen the second derivative $S^{^{\prime \prime }}<0$ is negative. The derivative 
\begin{equation}
\frac{\partial g(u,v,z)}{\partial z}=-\frac{2u(v-\frac{1}{v})\sin (z)}{
\left( v+\frac{1}{v}-(v-\frac{1}{v})\cos (z)\right) ^{2}} 
\end{equation}
is negative in $z\in [0,\pi ]$, and $f(u,v,z)=$ is positive in $
z\in [0,\pi ]$ (as can be seen from (\ref{gv2})). Then 
\begin{equation}
\frac{\partial h(u,v)}{\partial v}\leq 0\text{.} 
\end{equation}
That is, the inequality (\ref{utu}) improves as we make $v$ smaller keeping $u$ fixed. Therefore the best inequality holds for $v\rightarrow 1$ which gives $S^{^{\prime
\prime }}\leq 0$.

\section{Spheres in d+1 dimensions}
In this section we attempt a naive generalization of the above construction in more dimensions and find some difficulties related to the more complicated structure of the UV divergences in the entropy. 

It is not possible in general to divide the solid angle of the unit sphere in dimension $d>2$ with points located symmetrically for any $N$. This seems not to be a problem as long as we consider a large number of points with a constant density per unit solid angle. 
 Again our boosted sphere $D$ with boundary on the light cone has radius $\sqrt{Rr}$. It lies in the hyperplane given by the equation (\ref{plano}) and has center at the point
\begin{equation}
x_{1}=\frac{(R-r)}{2},\,\,\,\,x_{2}=...=x_{d\text{ }}=0\,,\,\,\,\,\,\,\,\,
\,\,t=\frac{(R+r)}{2} \,.
\end{equation}
The equation of $D$ is 
\begin{equation}
Rr=\left( x_1-\frac{R-r}{2}\right) ^{2}+x_{2}^2+...+x_{d}^2-\left( t-\frac{R+r}{2}\right)
^{2} \,, \label{circe}
\end{equation}
supplemented by the equation (\ref{plano}).

Each wiggly sphere resulting of a set in (\ref{ecu}) must have one of its vertices on $D$. Then we focus attention on the surface of $D$. One of this sets is the union of all the intersections of, let say $k$ rotated spheres.  Hence, at this level $k$ the radius of the wiggly sphere is given by the maximum radius of a point belonging to $k$ different spheres at the same time. In order to find this radius we have to take these $k$ spheres as much tightly packed in solid angle around the direction of $D$ as possible.  

As we increase $k$ we have to open up the solid angle around the direction of $D$. We have 
$dk=\frac{N}{\textrm{vol}(s^{d-1})}\, d\Omega$, where $\textrm{vol}(s^{d-1})=2 \pi^{\frac{d}{2}}/\Gamma(\frac{d}{2})$ is the total solid angle corresponding to the volume of the $d-1$ dimensional unit sphere. We can relate the differential of solid angle with the radius $l$ of the wiggly sphere (centered at the origin) by writing $d\Omega =\frac{d\Omega }{dA}\frac{dA}{dl}dl$
where $dA$ is the differential of area on the surface $D$. 
Since $D$ is a null surface, $dA=dA^{\perp }$, where this last differential is the area of the surface differential projected along  the null rays generating the light cone. On the other hand $dA^{\perp }$ is equal to $d\Omega \,l^{d-1}$.

Thus, we have
\begin{equation}
dk=\frac{N}{\textrm{vol}(s^{d-1})}\frac{1}{l^{d-1}}\frac{dA}{dl}dl \,.
\end{equation}
The  $dA$ is given in terms of the differential of azimuthal angle $d\theta$ on the sphere $D$ by
\begin{equation}
dA=\textrm{vol}(s^{d-2})(x^{\intercal})^{d-2}\,\sqrt{Rr}\, d\theta \,,
\end{equation}
where $x^{\intercal }=\sqrt{x_2^2+...+x_d^2}$ is the radius of the $d-1$ dimensional sphere which is the solution with fixed $t$ of the equations (\ref{plano}) and (\ref{circe}).  Note that as these points are on the cone we have $t=l$. We get from these equations
\begin{equation}
x^{\intercal }=2\frac{\sqrt{rR(l-r)(R-l)}}{(R-r)}\,. 
\end{equation}
The quantity $\frac{d\theta }{dl}$ is obtained setting $x_{3}=...=x_{d}=0$ on $D$ 
 and  computing
\begin{equation}
\frac{d\theta }{dl}=\frac{d\theta }{dt}=\left| \frac{d(x_{1}(t),x_{2}(t),t)}{\sqrt{Rr}dt}\right|
=\frac{1}{\sqrt{(R-l)(l-r)}} \,.
\end{equation}
Combining all together we obtain a relation giving the number of wiggly spheres in terms of the radius $l$,  
\begin{equation}
dk=N\frac{\textrm{vol}(s^{d-2})}{\textrm{vol}(s^{d-1})}\frac{(x^{\intercal })^{d-2}\sqrt{Rr
}}{l^{d-1}}\frac{dl}{\sqrt{(R-l)(l-r)}}\,. 
\end{equation}

At this point we are prepared to write down the inequality (\ref{ecu}) in the limit $N\rightarrow \infty$. Let us naively assume again that the microscopic 
structure of the wiggles does not contribute in the limit $N\rightarrow \infty$ and assimilate the entropy of these surfaces to entropy of spheres. We have 
\begin{equation}
S(\sqrt{Rr})\geq \frac{2^{d-2}\Gamma(d/2)}{\sqrt{\pi}\Gamma ((d-1)/2)}\int_{r}^{R}dl\,\frac{
\left( Rr\right) ^{\frac{(d-1)}{2}}\left( (l-r)(R-l)\right) ^{\frac{d-3}{2}}}{
\left( R-r\right) ^{d-2}l^{d-1}}S(l)\,.\label{misi}
\end{equation}

This inequality is an equation for the area term, $S(l)\sim l^{d-1}$ and the constant term $S(l)\sim \textrm{const}$. Setting $R=r+\varepsilon$ and expanding for small $\varepsilon$ we get
\begin{equation}
r S^{\prime\prime}(r)-(d-2) S^{\prime}(r)\le 0\,.\label{refi}
\end{equation}
A related inequality was obtained in \cite{further1} for the strip geometry.

Let us check inequality (\ref{refi}) in $d=3$ ($3+1$ dimensional space-time) for a conformal theory. On general grounds we expect an entanglement entropy of the form 
\begin{equation}
S(r)=c_{2} \, r^{2}+ c_{1} \,r+ c_{\log}\, \log(R) +c_0  \,,
\end{equation} 
with constant $c_2$, $c_1$, $c_{\log}$ and $c_0$.  
The relation (\ref{refi}) gives 
\begin{equation}
c_1+2 \frac{c_{\log}}{r}\ge 0\,.
\end{equation}
However, in $d=3$ the coefficient of the logarithmic term $c_{\log}$ is always negative \cite{solo,hhm}. Thus this inequality does not 
hold for small enough $r$. One could argue that $c_1$ has to compensate this up to the scale of the cutoff. However, if there is a covariant
 regularization of the entropy, one does expect it is possible to set $c_1=0$. This is because by dimensional arguments it must be $c_1\sim \epsilon^{-1}$, 
and this ultraviolet divergent term must be generated locally on the surface of the sphere. But there are no local geometric invariant quantities which 
integrated on the surface would give a dependence proportional to $r$ on the entropy \cite{masivocurva,lolo}. 

One could still argue that the inequality (\ref{refi}) has to hold for the leading divergent term in the entropy.  Thus, it follows from (\ref{refi}) that
 the function $(d-1)^{-1}r^{-(d-2)} S^{\prime}(r)$ decreases from the ultraviolet to the infrared, while at the fixed points it is simply $c_{d-1}$, the 
coefficient of the area term (at leading order in the cutoff). Then we have
\begin{equation}
\Delta c_2=c^{\textrm{uv}}_{d-1}- c^{\textrm{ir}}_{d-1}=-\int_0^\infty dr\, \left(\frac{S^{\prime}(r)}{(d-1)r^{d-2}}\right)^{\prime}\ge 0 \,.\label{fff}
\end{equation}
 However, it is expected that the leading divergent term is a property of the ultraviolet fixed point, and does not run with the RG. For the case of a massive scalar the renormalization of the area term is \cite{massiveterm}
\begin{eqnarray}
\Delta c_2&=&\gamma_d  \,\textrm{vol}(s^{d-1})\,m^{d-1}\,\log(m\epsilon) \,\hspace{1cm}\textrm{for} \,\,\,d \,\,\,\textrm{odd}\,,\nonumber\\
\Delta c_2&=&\gamma_d\,\textrm{vol}(s^{d-1})\, m^{d-1}\,\hspace{2.5cm}  \textrm{for}\,\,\, d \,\,\,\textrm{even}\,,\label{corr}
\end{eqnarray}
with $\gamma_d=(-1)^{\frac{d-1}{2}}[6 (4 \pi)^{\frac{d-1}{2}}((d-1)/2)!]^{-1}$ for $d$ odd, 
and $\gamma_d=(-1)^{d/2+1}[12 (2 \pi)^{\frac{(d-2)}{2}}(d-1)!!]^{-1}$ for $d$ even.  Thus, in these free examples, the change of the coefficient of the area  
term is subleading in powers of $\epsilon$ with respect to the leading term $\sim \epsilon^{-(d-1)}$. This invalidates the argument in this case since we know our inequality has to
 be corrected for subleading terms. Hence, it is not surprising that the corrections (\ref{corr}) can have any sign depending on the dimension.  

The reason for these subtleties with the inequality (\ref{refi}) has to reside in the replacement of the wiggly sphere entropy $\tilde{S}(r)$ by the one corresponding to a smooth sphere $S(r)$. 
For dimensions $d>2$ there should be different contributions due to the microscopic structure of the surface. For the case $d=3$ we can see at least two contributions. 
One is due to the different curvatures of the small patches of spheres with radius $\sqrt{r R}$ which cover the sphere of radius $l$. This has to give place to different 
logarithmically divergent terms in the entropies. Also, the trihedral angles might give a logarithmic contribution to the wiggly sphere which
 is not present in the smooth sphere. Some contributions to the entropy of singular entangling surfaces in four and higher dimensions have been recently studied \cite{xxx}.  
 In the worst case these missing terms could make the inequalities trivial. 
 These contributions are not present in the $d=2$ case, where there is no curvature for the arc patches, and the vertices have 
angle $\pi$. 

One way to convince oneself of the different behavior of the trihedral and planar angles is to think in an approximation of a sphere by 
plane polyhedrons with many phases. In the two dimensional case we have $N$ angles on the polygon boundary which deviate from $\pi$ by a quantity of order $N^{-1}$. 
As the logarithmic contribution for angles $\pi-\delta$ goes like $\delta^2$ for small $\delta$, the total logarithmic contribution is of order $N/N^2=N^{-1}$ and 
vanishes in the limit of a circle. This is another way to see the circle does not have logarithmic contributions. In contrast, the logarithmic contribution of the 
polyhedron must 
come from the trihedral vertices, and must sum up the non zero logarithmic term of the sphere in the limit.

\section{Final remarks}

The result of this paper for the monotonous RG running of the constant term of the circle entropy gives a QFT proof of the recent holographic
 result for this same quantity and the related ones for the free energy on a three sphere. The only principles underlying this property are  
relativistic covariance and unitarity.  
It would be an analogous in three dimensions to the two-dimensional Zamolodchikov's c-theorem if the fixed point value of the monotonous quantity could
 be intrinsically determined. Lacking of a good definition of $c_0$ at the conformal points we do not knot at the moment if $c_0$ can be globaly defined or it keeps a memory of the 
RG trajectory in interacting theories. 

One alternative which deserves future investigation is the possible definition of $c_0$ through the mutual information. The mutual information of two 
regions $I(A,B)=S(A)+S(B)-S(AB)$ is well defined in the continuum limit, and for finite systems with pure global state we have $I(A,-A)=2 S(A)$. Then, 
in quantum field theory we expect the mutual information between a sphere and the outside region separated by a distance $\epsilon$, $-A_{\epsilon}$, has 
an expansion $I(A,-A_{\epsilon})\sim \frac{k}{\epsilon}+2 \,\tilde{c}_0+{\cal{O}(\epsilon)}$ for small $\epsilon$. Here $\tilde{c}_0$ would be the 
regularization independent definition of $c_0$. However, at the moment it is not known how to extend the geometric construction to prove monotonicity 
for $\tilde{c}_0$ in the mutual information.  

Our result in two spatial dimensions also shows the coefficient of the area term on the entropy of circles is decreasing with the radius. The coefficient of the area term for a 
conformal theory must be a local property independent of the shape of the region. This suggests an interesting (albeit speculative)  interpretation for the decreasing of the coefficient of the area term. It could in principle be converted to a c-theorem like result by going outside the domain of QFT. There is in fact an intrinsic way to determine the area term in a theory 
which is coupled with gravity. In that case, the quantity $c_{1}$, the coefficient of the area term, is supposed to be related to the Newton constant through 
the Hawking entropy \cite{sasa} as 
\begin{equation}
c_{1}= \frac{\pi}{2 G }\,,
\end{equation}
where $G$ is the Newton constant. 
In this context, the meaning of (\ref{fff}) is that the Newton constant increases from the UV to the IR (the Planck mass decreases). When massive modes 
decouple their degrees of freedom are not computed any more in the IR theory but their 
contribution is remembered by the Newton constant through virtual loops. The Sakharov's induced gravity is the case in which gravity is free at the UV point and at 
the infrared $G$ is entirely generated by the integrated fields.         
 
The issue of divergences in the entanglement entropy in higher dimension is more subtle. We need a covariant regularization of the entropy which respect 
the SSA property. In order for this regularized entropy to be non trivial, that is, not given exclusively as boundary terms which cancel in the mutual 
information, there must be regions for which it takes negative values \cite{geo}. This situation is odd for the entropy in ordinary quantum theory of 
finite number of degrees of freedom, but it is natural for the classical entropy of fields. Nevertheless, this covariant entropy would still be ambiguous under 
redefinitions by addition of boundary terms such as the area term or the Euler number of the region \cite{geo}.

One interesting point which deserves future investigation is to find the missing terms in eq. (\ref{misi}). These terms must come from the 
microscopic structure of the surface. In particular, in $d=3$ we have to understand the logarithmic contributions of trihedral angles. We hope an improved inequality    
 could have a bearing on the monotonous behavior for 
the coefficient of the logarithmic term of the sphere, which is proportional to the Euler trace anomaly, 
and the best candidate for a c-theorem in odd spatial dimensions $d\ge 3$.

\section*{Acknowledgments}
This work was supported by CONICET, ANCyT, Universidad Nacional de Cuyo, and CNEA, Argentina. We were benefitted by discussions with Igor Klebanov, Hong Liu, Juan Maldacena and Rob Myers for improvements on the second version of this preprint.

\end{document}